 \documentstyle[aps,prd,epsfig,preprint,floats]{revtex} %single-column/doublespacing
\tightenlines	%%% requested by Los Alamos server (to save the disk space)
\input epsf.tex
\def\DESepsf(#1 width #2){\epsfxsize=#2 \epsfbox{#1}}
\newcommand{\ifm}[1]{\relax\ifmmode #1\else $#1$\fi}
 
\newcommand{\beq   }{\begin{equation}}
\newcommand{\eeq   }{\end{equation}}
\newcommand{\beqn  }{\begin{eqnarray}}
\newcommand{\eeqn  }{\end{eqnarray}}
\newcommand{\bi    }{\begin{itemize}}
\newcommand{\ei    }{\end{itemize}}
\newcommand{\bc    }{\begin{center}}
\newcommand{\ec    }{\end{center}}
\newcommand{\bd    }{\begin{description}}
\newcommand{\ed    }{\end{description}}
\newcommand{\bHuge }{\begin{Huge}}
\newcommand{\bhuge }{\begin{huge}}
\newcommand{\bLARGE}{\begin{LARGE}}
\newcommand{\bLarge}{\begin{Large}}
\newcommand{\blarge}{\begin{large}}
\newcommand{\eHuge }{\end{Huge}}
\newcommand{\ehuge }{\end{huge}}
\newcommand{\eLARGE}{\end{LARGE}}
\newcommand{\eLarge}{\end{Large}}
\newcommand{\elarge}{\end{large}}
\def \mc {\multicolumn}

\def \gtsim    {\relax\ifmmode{\mathrel{\mathpalette\oversim >}}
                  \else{$\mathrel{\mathpalette\oversim >}$}\fi}
\def \ltsim    {\relax\ifmmode{\mathrel{\mathpalette\oversim <}}
                  \else{$\mathrel{\mathpalette\oversim <}$}\fi}
\def\oversim#1#2{\lower4pt\vbox{\baselineskip0pt \lineskip1.5pt
            \ialign{$\mathsurround=0pt#1\hfil##\hfil$\crcr#2\crcr\sim\crcr}}}
\newcommand{\gev}  {\mbox{${\rm GeV}$}}

\newcommand{\pgev} {\mbox{${\rm GeV}/c$}}
\newcommand{\gevcc}{\mbox{${\rm GeV}/c^2$}}

\newcommand{\mgev} {\mbox{${\rm GeV}/c^2$}}

\newcommand{\mtev} {\mbox{${\rm TeV}/c^2$}}

\newcommand{\invpb}{\mbox{${\rm pb}^{-1}$}}

\newcommand{\invfb}{\mbox{${\rm fb}^{-1}$}}

\newcommand{\intlum}{\mbox{${ \int {\cal L} \; dt}$}}
\newcommand{\pt}  {\mbox{$p_{T}$}}

\newcommand{\mets}{\mbox{${E\!\!\!\!/_T}$}}
\newcommand{\met} {\mbox{${E\!\!\!\!/_T}$}}

\newcommand{\bbar} {\mbox{$\overline{b}$}}

\newcommand{\ppbar}{\mbox{$p\overline{p}$}}

\newcommand{\ttbar}{\mbox{$t\overline{t}$}}

\newcommand{\epem} {\mbox{$e^+e^-$}}
\newcommand{\mpmm} {\mbox{$\mu^+\mu^-$}}
\newcommand{\azero}{\ifm{A_0}}
\newcommand{\tanb}{\ifm{\tan\beta}}
\newcommand{\mzero}{\ifm{m_0}}
\newcommand{\mhalf}{\ifm{m_{1/2}}}
\newcommand{ \gravitino}{\mbox{$\tilde{G}$}}
\newcommand{ \gluino}   {\mbox{$\tilde{g}$}}
\newcommand{ \squark}   {\mbox{$\tilde{q}$}}

\newcommand{ \squarkb}  {\mbox{$\bar{\tilde{q}}$}}
\newcommand{ \slepton}  {\mbox{$\tilde{\ell}$}}
\newcommand{ \sleptonR}  {\mbox{$\tilde{\ell}_{R}$}}

\newcommand{ \stauone}  {\mbox{$\tilde{\tau}_{1}$}}

\newcommand{ \stau}     {\mbox{$\tilde{\tau}$}}

\newcommand{ \snu}      {\mbox{$\tilde{\nu}$}}

\newcommand{ \sbottomone}{\mbox{$\tilde{b}_{1}$}}
\newcommand{ \sbottomoneb}{\mbox{$\bar{\tilde{b}}_{1}$}}
\newcommand{ \sstop}    {\mbox{$\tilde{t}$}}

\newcommand{ \stopone}  {\mbox{$\tilde{t}_{1}$}}
\newcommand{ \stoponeb} {\mbox{$\bar{\tilde{t}}_{1}$}}

\newcommand{ \schi }    {\mbox{$\tilde{\chi}$}}
\newcommand{ \lsp}      {\mbox{$\tilde{\chi}_{1}^{0}$}}
\newcommand{ \schionezero }{\mbox{$\tilde{\chi}_{1}^{0}$}}
\newcommand{ \schitwozero }{\mbox{$\tilde{\chi}_{2}^{0}$}}

\newcommand{ \schionepm }{\mbox{$\tilde{\chi}_{1}^{\pm}$}}
\newcommand{ \schionep } {\mbox{$\tilde{\chi}_{1}^{+}$}}
\newcommand{ \schionem } {\mbox{$\tilde{\chi}_{1}^{-}$}}

\newcommand{ \bsmumu }{\mbox{$B_{s} \to \mu^+ \mu^-$}}
\newcommand{ \bsgam }{\mbox{$b \to s \gamma$}}
\def \PRL      {Phys. Rev. Lett.~}

\def \PRD      {Phys. Rev. D}

\def \PLB      {Phys. Lett. B}
\def \NPB      {Nucl. Phys. B}

\def \etal     {\relax\ifmmode{et \; al.}\else{$et \; al.$}\fi}
\newcommand{\ie}{$i.e.$}
\newcommand{\eg}{$e.g.$}
\newcommand{\eps}{\mbox{$\epsilon$}}
\def \calR     {\relax\ifmmode{{\cal R}}\else{${\cal R}$}\fi}
\def \Dzero    {\relax\ifmmode{{\rm D\O}}\else{D\O}\fi}
\def \DzeroC   {\relax\ifmmode{{\rm D\O\ Collaboration}}
	\else{D\O\ Collaboration}\fi}
\def \CDFC   {\relax\ifmmode{{\rm CDF Collaboration}}
	\else{CDF Collaboration}\fi}
\def \CDFIIC   {\relax\ifmmode{{\rm CDF II Collaboration}}
	\else{CDF II Collaboration}\fi}
\begin{document}
\pagestyle{empty}
%%%\draft
%%%222 \twocolumn[\hsize\textwidth\columnwidth\hsize\csname
%%%222 @twocolumnfalse\endcsname
\preprint{\hbox{CDF/PUB/EXOTIC/PUBLIC/6196}}
\title{
\boldmath  Prospects of Discovery for Supersymmetry at the Tevatron}
\author{Teruki Kamon\footnote{Representing the CDF and D\O\ Collaborations. 
Plenary talk at 10th  International Conference on Supersymmetry
and Unification of Fundamental Interactions (SUSY02),
June 17-23, 2002, DESY, Hamburg, Germany.}\\
{\it e-mail: t-kamon@tamu.edu} \\[.2in]
Department of Physics, Texas A\&M University,  
College Station TX 77843-4242, USA}
%\date{\today}
%
\maketitle
%%% Need this line to suppress the page number 1 on the 1st page!?!?!
\thispagestyle{empty}
%%%

%%%%%%%%%%%%%%%%%%%%%%%%%%%%%%%%%%%%%%%%%%%%%%%%%%%%
%%%222 \twocolumn[\hsize\textwidth\columnwidth\hsize\csname
%%%222 @twocolumnfalse\endcsname

\begin{abstract} 
We summarize a discovery potential for supersymmetric particles
at the $\ppbar$ collider of Tevatron 
with center-of-mass energy $\sqrt{s}$ = 2 TeV and 
integrated luminosity $\intlum$ = 15-30 \invfb. 
Any direct search is kinematically limited 
to below 450 \mgev.
We, however, have a unique opportunity to test various supersymmetric 
scenarios by
a measurement of the branching ratio for the rare decay mode
\bsmumu.
Using the background estimate in the CDF analysis of \bsmumu\ in Run I,
we investigate the prospects for studying this mode in Run II.
CDF would be sensitive to this decay for a branching ratio 
$> 1.2  \times 10^{-8}$ with 15 \invfb\
 (or, if a similar analysis holds for \Dzero,
$>6.5\times 10^{-9}$ for the combined data). 
For $\tanb > 30$, the \bsmumu\ search can probe
the SUSY parameter space that cannot be probed 
by direct production of SUSY particles at  Run II.
An observation of  \bsmumu\ with 
a large branching ratio $> 7(14) \times 10^{-8}$ 
 (feasible with only  2 \invfb) 
would be sufficient to exclude the mSUGRA model for
$\tan\beta \leq 50 (55)$
including other experimental constraints.
For some models, 
the branching ratio can be large  enough to be detected even
for small $\tan\beta$ and large \mhalf.\\
\\ %PACS numbers: 14.80.Ly,  12.60.-i, 12.60.Jv, 13.85.Rm, 11.30.Pb
\end{abstract}
%%%111 \pacs{ }   % for single-column and double space

%%%222 \vskip2pc]

%%%%% Main Text %%%%%%%%%%%%%%%%%%%%%%%%%%%%%%%%%%%%%%%%
\newpage
\section {Introduction} 

The Fermilab Tevatron collider
will define the high energy frontier of particle physics 
while CERN's Large Hadron Collider is being built.
The first stage (Run IIa) of the Tevatron collider Run II (March 2001-2005) 
will deliver at least 2 \invfb\  of integrated luminosity per experiment 
at $\sqrt{s}$ = 2 TeV.
Major upgrades of the CDF and \Dzero\ detectors \cite{CDF2det_tdr,D02det_tdr}
have been completed and
we have beeing taking data.
In the second stage (Run IIb) of Run II (starting in 2006),
we expect  15 \invfb\  of integrated luminosity per experiment at 2 TeV.
Further upgrades of the two detectors are been undertaken.

Among various features, the detectors have the ability
to trigger on displaced vertices from bottom and charm decays
using a precise microvertex detector to enchance the Higgs search and
the physics with top quarks.
Searches for supersymmetry (SUSY) are among the 
main priorities 
along with Higgs and top physics for Run II.

Supersymmetry uniquely opens the possibility to directly connect 
the Standard Model (SM)
with an ultimate unification of the fundamental interactions.
With the results on electrweak and strong gauge couplings 
from CERN's \epem\ collider LEP experiments and
the top quatk mass at CDF and D\O\ at the Tevatron, 
the models of SUSY have become more predictive and
require a spectrum of new particles below a few \mtev.
Thus SUSY  represents  a  natural candidate for the new physics expected 
to occur in the TeV energy  domain. 

In this paper, we summarize the prospects of SUSY searches/discovery
either directly through collider processes or
indirectly through rare processes at the Tevatron.

\section{SUSY Models}

One of the difficulties in determining predictions of
generic Minimal Supersymmetric Standard Model (MSSM) 
lies in the large number of new parameters 
(over 100 free parameters) the theory implies. 
One may consider a theoretical framework to reduce the number of
free parameters.

Fortunately SUSY models apply to a large  number of different
accelerator and cosmological phenomena, 
and a great  deal of effort has been
involved in recent years to use the data to  limit the parameter space. 
Part of the difficulty in doing this resides 
in  the success of the model in not
disturbing the excellent agreement of the  precision tests of the SM 
\cite{Langacker:2001ij} due to
the SUSY decoupling theorems which  suppress SUSY contributions 
at low energies.
Historically, the absence of  flavor changing neutral currents 
at the tree level played an important role  in the construction of the SM. 
They represent therefore an important class  of phenomena 
that might show the presence of new physics, 
since the SM and  the SUSY contributions contribute first at the loop
level with comparable  size.  
Thus the decay \bsgam\ has been a powerful 
tool in limiting  the SUSY parameter space.

Extensive Monte Carlo (MC) studies were carried out during 1998
on the following four topics
to maximize the direct SUSY/Higgs searches in Run II:
(i)~supergravity (SUGRA)~\cite{shw_sugra}, 
(ii)~Gauge-mediated SUSY Breaking (GMSB)~\cite{shw_gmsb},
(iii)~beyond the MSSM~\cite{shw_btmssm}, 
and (iv)~Higgs~\cite{shw_higgs}.
The readers can refer to Ref.~\cite{tev_susy_review}
for a summary of Run~I SUSY searches.
Among the experimental aspects, we conclude that
it is important to have excellent triggering/tagging and identification
for $b$'s, $\tau$'s, $\gamma$'s as well as $e$'s and $\mu$'s.
Thus we develop (A) low \pt\ lepton+track trigger, where
the ``track'' object can be electron, muon, or 
hadronically decaying $\tau$-lepton,
(B1) $\geq$2-jets trigger + \mets\ ($>25\ \gev$)
and 
(B2) $\geq$2-jets trigger + \mets\
($>20\ \gev$  with $b$/$c$ tagging), 
instead of inclusive \mets\ trigger,
(C) better trigger/identification for prompt/displaced photons.
Trigger B1 is extremely useful for
reliable parametrization of high \mets\ distribution due
to QCD events to reduce the systematic uncertainty \cite{shw_sugra}.

We first consider mininal SUGRA (mSUGRA) and GMSB frameworks
as examples of direct searches for SUSY production,
which characterize the experimental triggers and analyses
in $b$'s, $\tau$'s, $\gamma$'s.
Details of all SUSY models and experimental prospects
can be found in Refs.~\cite{shw_sugra,shw_gmsb,shw_btmssm}.
We then take 
the decay \bsmumu\
as an example of powerful tool in limiting 
SUSY parameter space at the Tevatron.

\section{Direct Searches}

\subsection{Testing mSUGRA}
\label{sec:ds_sugra}

The mSUGRA model \cite{Chamseddine:jx,sugra2} depends on four parameters:
\mzero\ (the universal scalar mass at $M_G$), 
\mhalf\ (the universal gaugino mass at $M_G$), 
\azero\  (the universal cubic soft breaking mass at $M_G$),  and 
$\tanb$ (the ratio of two SUSY Higgs 
vacumm expectation values at the electroweak scale).
In addition, the sign of $\mu$ (the  Higgs mixing parameter) is  arbitrary.  
With R parity invariance, 
the lightest neutralino ($\schionezero$) is assumed to be
the lightest suppersymmetric particle (LSP) and it is bino-like
and  stable.
The $\schionezero$ then would pass through the detector without interaction.
We fix \azero = 0 and the sign of $\mu$ to be positive for simplicity, 
otherwise stated.
Here the ISAJET sign convention for $\mu$ is used.

\subsubsection{Chargino-Neutralino Associated Production}

The trilepton signal arises when
both the lightest chargino  ($\schionepm$) and 
the next-to-lightest neutralino ($\schitwozero$) decay
leptonically in $\ppbar \to  \schionepm \schitwozero + X$.
An initial study of the final state of trilepton ($\ell\ell\ell$) plus \mets\
for high luminosity ($\gtsim$10 \invfb) at the Tevatron
was made in Ref. \cite{Kamon:1994}
for direct $\schionepm \schitwozero$ production.
Here $\ell\ell\ell = eee, ee\mu, e\mu\mu, \mu\mu\mu$
by excluding electron or muon from tau ($\tau$) lepton leptonic decay.
Further studies were also made for
$\ppbar \to  3\ell + X$,
including all SUSY production processes (\eg, $\slepton\snu$)
and its decays
\cite{Baer:1998,Barger:1998hp,Matchev:2000,shw_sugra,Dedes:2002}.
Electron or muon from the leptonic decay mode of  $\tau$ lepton were
accepted.
This requires CDF and D\O\ experiments
to trigger and identify the leptons 
with low \pt\ ($\gtsim 10\ \pgev$).
They also improved the SM background calculations
including effects from $W^*$, $Z^*$, and $\gamma^*$.

\begin{figure}[t]
\centerline{\hbox{\psfig{figure=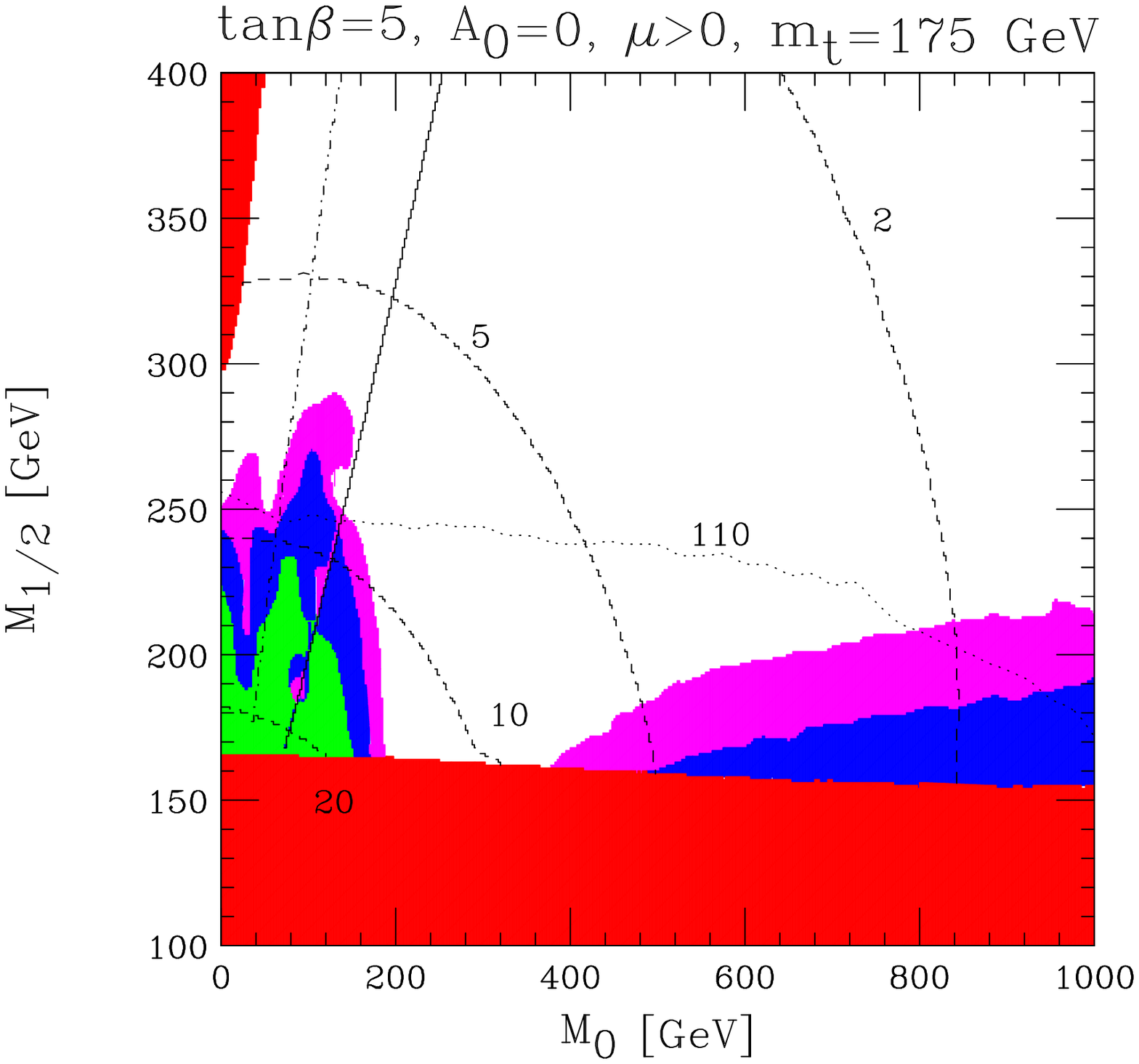,height=3in,angle=90}}
{\psfig{figure=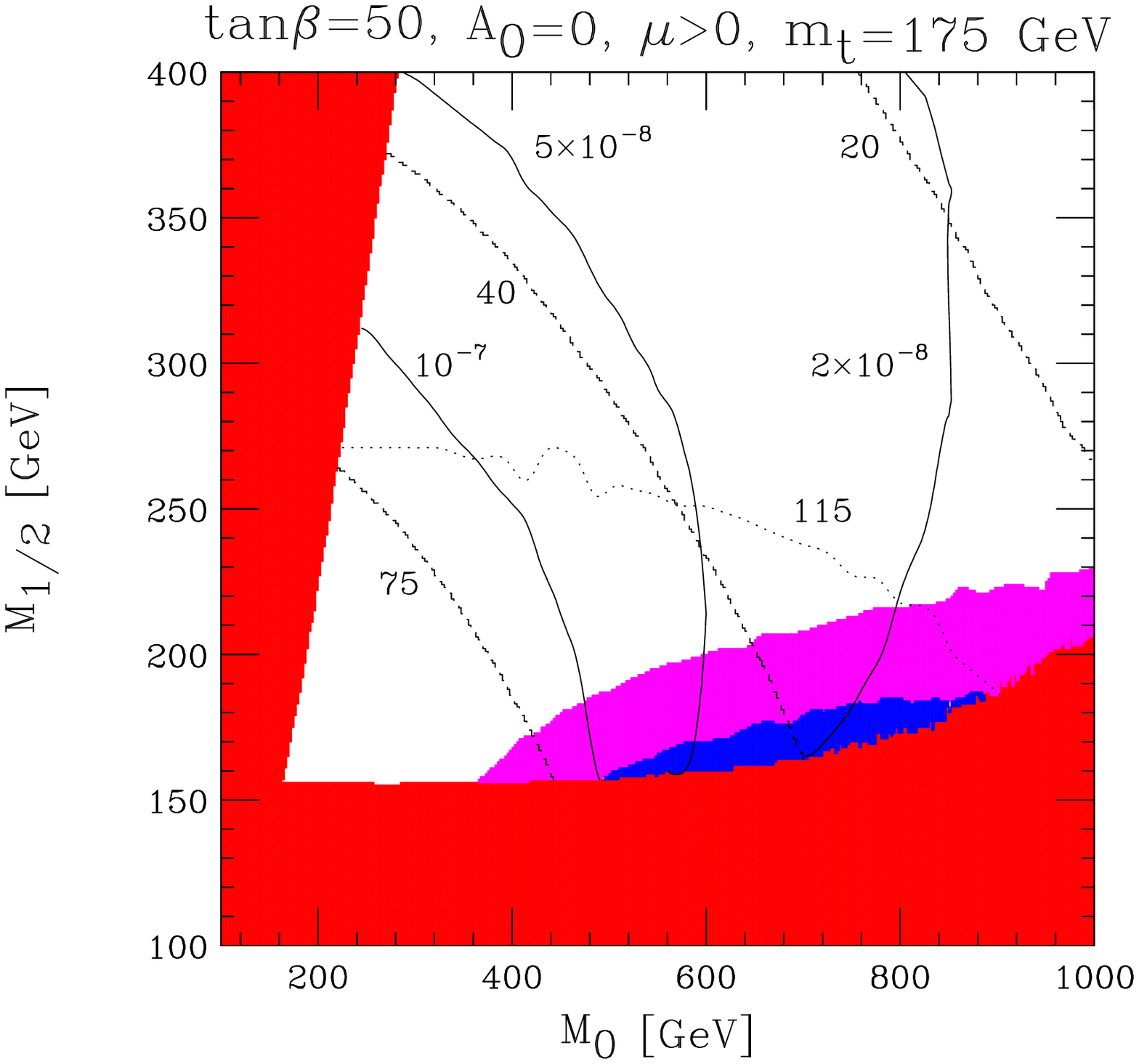,height=3in,angle=90}}}
\caption{Regions of the  $\mhalf$-$\mzero$ plane
where the trilepton events should be detectable at a level of
5$\sigma$ significance
for  $\tan\beta=5$ (left) and  $\tan\beta=50$ (right)
  \protect\cite{Dedes:2002}. 
Three areas are shown
for an integrated luminosity of 30 \invfb\ (magenta),
10 \invfb\  (blue) and 2 \invfb\ (green) and from top to bottom,
respectively. 
The large red regions
are excluded by theory and experiment.
Dashed lines represent the SUSY contribution to the
muon anomalous magnetic moment (in units of $10^{-10}$) and
the dotted lines  are iso-mass contours of the lightest neutral Higgs
boson.
The solid contour (only for $\tan\beta=50$) indicates the
prediction for the branching ratio $Br[\bsmumu]$.  
In the left
($\tan\beta=5$) plot the solid line indicates where
$M_{\tilde{\chi}^\pm_1}=M_{\tilde{\tau}_1}$ and 
the dot-dashed line
$M_{\tilde{\chi}^\pm_1}=M_{\tilde{\nu}_\tau}$.}
\label{fig:Dedes2002_3L}
\end{figure}

The studies
in Refs.~\cite{Baer:1998,Matchev:2000,shw_sugra,Dedes:2002} 
have included hadronically decaying $\tau$ lepton as well. 
This requires both experiments
to trigger and identify the $\tau$ leptons 
with low \pt\ ($\gtsim 10\ \pgev$).
With high luminosity in Run II, an inclusive lepton trigger becomes more
difficult because of such a large trigger rate.
Thus a generic dilepton trigger, namely lepton + track trigger, is
neccessary \cite{shw_sugra}.

In those studies, the final states of $3\ell + \mets$,
$\ell^{\pm}\ell^{\pm} + \tau_h + \mets$ are found to be
the best channels for the study of
chargino-neutralino associated production.
Trigger A, mentioned earlier, will play a key role to maximize
the experimental sensitivity in this channels.
Figure \ref{fig:Dedes2002_3L} shows $5\sigma$ discovery
reach in the trilepton  channel
in mSUGRA for small \tanb\ = 5 and large \tanb\ = 50.
We will be sensitive upto $\mhalf\ \simeq 250\ \mgev$ 
if $\mzero\ \ltsim\ 200\ \mgev$ and low $\tanb$
(\eg, 5).
With large $\tanb$ (\eg, 50), 
upto $\mhalf~\simeq~200~\mgev$ 
if $\mzero\ \gtsim\ 500\ \mgev$. 
It should be noted that 
the CDF and D\O\ analyses \cite{CDF3L:1998,D03L:1998}
of the trilepton channel in Run I (about 100 \invpb) 
were limited to $\tanb = 2$, $\mu < 0$ 
within the MSSM framework.
The chargino mass
upto about 80 \mgev\ 
($\mhalf\ = 75\ \mgev$) was excluded 
for $\tanb = 2$, $\mu < 0$ and $\azero = 0$
if $M_{\squark} = M_{\gluino}$ within mSUGRA  \cite{Kamon:1998}.

\subsubsection{Gluinos and Squarks Production}

Gluinos  ($\gluino$) and squarks ($\squark$) are pair-produced at the Tevatron.
Within the mSUGRA framework, $M_{\squark}\ \gtsim\ 0.85\ M_{\gluino}$.
Thus, there appear two representative parameter regions in terms of
the production:
(i)~$\gluino\gluino, \gluino\squark$, and $\squark\squarkb$ productions where
	$M_{\squark} \simeq M_{\gluino}$ and
(ii)~$\gluino\gluino$ production  where $M_{\squark} \gg M_{\gluino}$.

In most of the parameter space accessible at the Tevatron,
the left-chiral squark
dominantly decays into a quark and either a $\schionepm$ or a $\schitwozero$.
Those branching ratios are
$B(\squark_{L} \to q^{\prime} \schionepm) \simeq 65\%$ and
$B(\squark_{L} \to q \schitwozero) \simeq 30\%$.
Since $M_{\squark} - M_{\schionepm} > M_{\schionepm} - M_{\lsp}$,
the jet in the $\squark_{L} \to q^{\prime} \schionepm$ 
(or $\squark_{L} \to q \schitwozero$) decay likely has larger
$E_T$ than those in the $\schionepm \to q \bar{q}^{\prime} \lsp$ 
(or $\schitwozero \to q q^{\prime} \lsp$) decay \cite{Note1}.
Similarly, at least one jet in the gluino decay 
($\gluino \to q \bar{q}^{\prime} \schionepm$ or 
$q \bar{q} \schitwozero$ through a real of virtual squark) has large $E_T$.
Thus, pair-produced squarks and gluinos have at least two large-$E_T$ jets
associated with large \met.
Furthermore,
the jet multiplicity tends to be larger for events
with gluino than with squark.
The final state with lepton(s) is possible
due to leptonic decays of
the $\schionepm$ and/or $\schitwozero$. 
The branching ratio to the final state with
two or more leptons strongly depends
on the value of $\tanb$.
This leads us to  look for
SUSY events with final states of from jets + \mets\ 
and 1$\ell$ + jets + \mets\ 
($\ell$ = $e$ or $\mu$) \cite{Krutelyov:2001}.
We restrict the parameter space
so that lighter third generation squarks ($\sbottomone$ and $\stopone$)
remain heavier than the  $\schionepm$ 
and the $\schitwozero$.

\begin{figure}
\begin{center}
    \epsfig{file=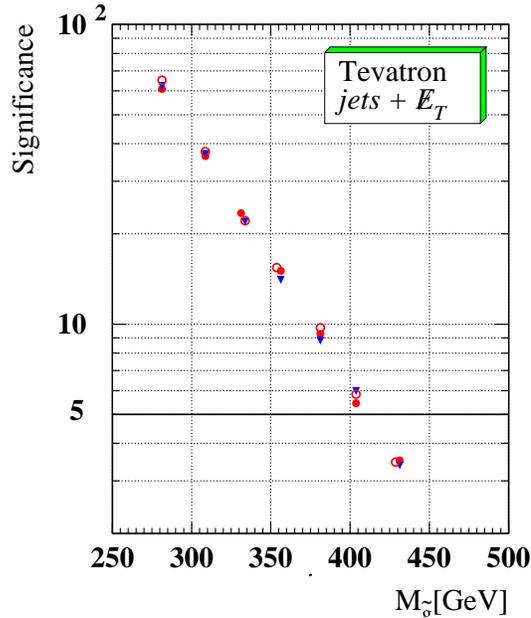,height=250pt}
    \caption{Significance as a function of $M_{\gluino}$ 
	($M_{\squark} \simeq M_{\gluino}$) for
	$\tan\beta = 3$ (filled circles), 10 (down triangles), and
	30 (open circles) in jets + \met\ channel (15 \invfb) 
	at the Tevatron~\protect\cite{Krutelyov:2001}.}
    \label{fig:tev_metjet_tanB3_10_30}
\end{center}
\end{figure}

In  the jets + \mets\ channel, for example,
an optimization of
cuts could be made on $N_j$, \met, and $M_{S_{2}}$ 
($\equiv \met\ + E_T^{j_1} + E_T^{j_2}$) \cite{Krutelyov:2001}.
The final selection cuts are:
(a)~$N_j \geq 4$;
(b)~veto on isolated leptons ($e$ or $\mu$) with $p_T > 15\ \pgev$;
(c)~$\met > 100\ \gev$;
(d)~$\Delta \phi_{j \mets} > 30^{\circ}$;
(e)~$M_{S_2} > 350\ \gev$.
The SM background sizes are estimated to be
25 fb for $\ttbar$ events,
38 fb for $W/Z$ + jets events,
1 fb for diboson process,
and 9 fb for QCD events,
totaling 73 fb.
Figure~\ref{fig:tev_metjet_tanB3_10_30} is the significance 
as a function of  $M_{\gluino}$ where $M_{\squark} \simeq M_{\gluino}$
at $\tan\beta$ = 3, 10, and 30 ($\mu > 0$ and $\azero = 0$).
The strongest reach in 5$\sigma$ significance
is 410 \mgev\ ($\mhalf\ \simeq\ 160\ \mgev$)
for 15 \invfb. 
There is no significant $\tanb$ dependence.
This can be compared to
280 \mgev\ for 100 \invpb,
360 \mgev\ for 2~\invfb, and
440 \mgev\ for 30 \invfb.
This channel has  huge QCD background events,
so that Trigger B1 will play a key role to minimize
systematic uncerainty in understanding the size
of the QCD events.

In the 1$\ell$ + jets + \mets\ channel, 
the gluino mass limits are less stringent than
those in the jets + \mets\ channel \cite{Krutelyov:2001},
except for smaller \mzero\ values.
It will be essential to combine limits from the two channels
to maximize the sensitivity in Run II.

It should be noted that 
the D\O\ and CDF analyses of the jets + \mets\ channel
within the mSUGRA framework ($\azero = 0$) in Run I
were limited to $\tanb = 3$, $\mu < 0$ \cite{D0metjets:1999,CDFmetjets:2002}.
The stringent lower limit on the gluino mass at 95\% C.L. is 300 \mgev\ 
($\mhalf\ \simeq  130\ \mgev$) 
for $\tanb = 3$ and $\mu < 0$ if $M_{\squark} = M_{\gluino}$
 \cite{CDFmetjets:2002}.
The D\O\  analysis
of the 1$e$ + jets + \mets\ channel in Run I
was also limited to $\tanb = 3$, $\mu < 0$ \cite{D0metjets1L:2002}
and 
the gluino mass limit was less stringent than
those in the jets + \mets\ channel.

\subsubsection{Stop and Sbottom Production}

A large mixing angle $\theta_{\sstop}$ between the superpartners of the
left-chiral and the right-chiral  stop quarks,
$\tilde t_L$ and $\tilde t_R$  respectively, 
form two squark mass eigenstates where $M_{\tilde t_1} < M_{\tilde t_2}$.
The $\stopone$ could substatially be lighter than other squarks. 
mSUGRA with smaller $\mzero$ and/or larger $|\azero|$
would have a light stop  in their spectrum.
In contrast, lighter sbottom  ($\sbottomone$) can appear only at
small $\mzero$ and small $\mhalf$ in the mSUGRA models.
In addition, they
are always accompanied by light stops, except $\tanb > 20$ and $\mu < 0$.

For general studies, 
we simply assume eithe $\sbottomone$  or $\stopone$ is
the lightest squark. 
A comprehensive study on prospects of those searches can be found 
in Refs.~\cite{Demina:2000,shw_sugra}. 
Decays studied for the $\stopone$ or $\sbottomone$ in Run II are:
(i)~$\sbottomone \to b \schionezero$,
(ii)~$\stopone \to c \schionezero$,
(iii)~$\stopone \to b l \snu (b \slepton \nu)$,
(iv)~$\stopone \to b \schionepm$, 
followed by $\schionepm \to W^{(\ast)} \schionezero \to \ell \nu \schionezero$,
(v)~$\stopone \to b W \schionezero$.

Table~\ref{tab:sugra_stopsbottom} is a summary of the
maximum sensitivity in the searches.
In Run II, we should also consider the case where
the braching ratio for
$\stopone \to b \schionepm^{(\ast)} \to b \stau \nu$ is
nearly 100\%. 
Trigger A should enhance the sensitivity of the serach.
It should be noted that the above studies are based on
bino-like LSP. 
For higgsino-like LSP,
the search strategy needs to be modified and 
its prospects can be found in Ref.~\cite{Demina:2000}.

\begin{table}
\caption{Discovery reaches
on $M_{\sbottomone}$ and  $M_{\stopone}$
expected in Run II~\protect\cite{Demina:2000}.
The Run I limits are after taking into account
LEP2 limits on $M_{\schionezero}$,
$M_{\slepton}$, and $M_{\schionepm}$.
In Run II, 
$\ell b j \mets$ and  $cc\mets$ final states will be
 explored by Trigger B2.}
%\vspace{0.2cm}
\begin{center}
\begin{tabular}{ l l  l  l l }
\hline 
 Decay &  Subsequent Decay
	& Final State of
 & \mc{2}{c}{Discovery Reach in $M_{\stopone}$ or $M_{\sbottomone}$} \\
 ($Br$ = 100\%) &	& 
  $\sbottomone \sbottomoneb$ or $\stopone \stoponeb$ & 
  @20 \invfb & (Run I)\\
\hline
$\sbottomone \to b \schionezero$ &
	& $bb \mets$
	& 260 \mgev & (146 \mgev\ \cite{CDFmetbb:2000})\\
$\stopone \to c \schionezero$ &
	& $cc \mets$
	& 220 \mgev & (116 \mgev\ \cite{CDFmetbb:2000})\\
$\stopone \to b l \snu$ & $\snu \to \nu \schionezero$ 
	& $\ell^+\ell^- b \mets$
	& 240 \mgev & (140 \mgev\ \cite{Pagliarone:2002}) \\
$\stopone \to b l \nu \schionezero$ & 
	& $\ell^+\ell^- b \mets$
	& - & (129 \mgev\ \cite{Pagliarone:2002}) \\
$\stopone \to b \schionepm$ & $\schionepm \to W^{(\ast)}  \schionezero$
	& $\ell b j\mets$ and $\ell^+\ell^- j \mets$
	& 210 \mgev & (-)\\
$\stopone \to b W \schionezero$  & 
	&  $\ell b j\mets$ 
	& 190 \mgev & (-) \\
\hline 
\end{tabular}
\end{center}
\label{tab:sugra_stopsbottom}
\end{table}

The LEP limits on $\schionepm$ and $\slepton$ masses \cite{LEPlimits}
leave the decays of
(a)~$\sbottomone \to b \schionezero$ \cite{CDFmetbb:2000,D0metbb:1999},
(b)~$\stopone \to c \schionezero$ \cite{CDFmetbb:2000},
(c)~$\stopone \to b \ell \snu$ (three-body decay, $M_{\snu} = M_{W}$)
$\to b \ell \nu \schionezero$ \cite{Pagliarone:2002},
and
(d)~$\stopone \to b \ell \nu \schionezero$ 
(four-body decay, $M_{\snu} = M_{\stopone} - M_{b}$) \cite{Pagliarone:2002}
in Run I.
A summary of the mass limits is also provided in
Table~\ref{tab:sugra_stopsbottom}.
For cases (c) and (d), we simply assumed 
the three final states of 
$b e \nu \schionezero$, 
$b \mu \snu \schionezero$,
and $b \tau \nu \schionezero$ have the same branching ratio of 33.3\%.

\subsection{Testing GMSB}

The GMSB models are
generally distinguished 
by the presence of a nearly massless Goldstino ($\gravitino$)
as the LSP. 
The next-to-lightest SUSY particle (NLSP) decays to its partner and 
the $\gravitino$. 
Depending on the SUSY breaking scale ($\sqrt{F}$) , 
these decays occur
promptly ($\sqrt{F}\ \ltsim$ a few 100 TeV) 
or on a scale comparable to the size of a collider detector
(a few 100 TeV $\ltsim\  \sqrt{F}\ \ltsim$ a few 1000 TeV).
For $\sqrt{F}$ much larger than a few 1000 TeV, 
the NLSP decay takes place well outside a collider detector
and are not directly relevant to accelerator physics.
Thus we consider a systematic analysis based on a classification
in terms of the identity of the NLSP
and its decay length within the minimal GMSB models.
The models can be specified in terms of six parameters~\cite{shw_gmsb}:
$N_{m}$ (the number of generations of messenger fields),
$M_{m}$ (an overall SUSY mass for the messengers),
$\Lambda$ (the effective visible sector SUSY breaking parameter 
$= F_{S}/M_{m}$),
$C_{G}$ (the ratio of the messenger sector SUSY breaking order parameter
to the intrinsic SUSY breaking order parameter,  $F/F_{S}$,
 controling the coupling to the Goldstino),
in addition to
$\tanb$ and sign($\mu$).
The NLSP decay length scales like $C_{G}^{2}$.

\begin{table}[t]
\caption{Discovery reach (5$\sigma$ significance) on SUSY mass for
various NLSP scenarios 
in minimal GMSB models~\protect\cite{shw_gmsb}. 
In $c\tau$ column, ``p'' and ``d'' indicate
prompt and displaced decays of NLSP, respectively, while
``ll'' for long-lived NLSP.
$\gamma_{d}$ indicates a displaced photon.
$\delta_{im}$ indicates an impact parameter.}
%\vspace{0.4cm}
\begin{center}
\begin{tabular}{ l l l l l l}
\hline 
 \mc{1}{c}{NLSP} & Decay Mode & $c\tau$ & Prod. & 
	\mc{1}{c}{Key Final State(s)} & Discovery Reach \\
	&	&	&	&	& @30 \invfb\\
\hline
Bino $\schionezero$ & $\gamma + \gravitino$ & p &
	all &
	$\gamma\gamma\mets+X$ & 
	 340 \mgev\ ($\schionepm$) \\
	& 	 & d &
	all &
	$\gamma_{d} jj \mets$ or $\gamma\gamma\mets+X$ & 
	 300 \mgev\ ($\schionepm$) \\
	& 	 & &
	 &
	 & 
	 ($c\tau$ = 50 cm) \\
\hline
Higgsino $\schionezero$ & 
	$(h, Z, \gamma) + \gravitino$ & p &
	all &
	$(hh, h\gamma, hZ, Z\gamma, ZZ, \gamma\gamma)\mets+X$ &
	220 \mgev\ ($\schionepm$)		\\
              &  & d &
	all &
	$\delta_{ip} < 0$ for $h \to bb$, $Z \to \ell^+ \ell^-$ &
	--		\\
              &  & d &
	all &
	$\gamma_{d} +X$ &
	-- \\
\hline 
$\stau$ & $\tau + \gravitino$ & p &
	all &
	$\ell\ell\ell j \mets$, $\ell^{\pm}\ell^{\pm} jj \mets$, 
		$\tau_h \tau_h \mets$ &
	230 \mgev\ ($\schionepm$)\\
 	&  & &
	&  &
	120 \mgev\ ($\stauone$)\\
	&  & ll &
	all &
	$\mu$($dE/dx$) + $\ell\ell(M_{\ell\ell}>150\ \mgev)$ &
	420 \mgev\ ($\schionepm$) \\
	&  &  &
 	& &
	210 \mgev\ ($\stau$)\\
	&  & ll &
	all &
	$\mu$($dE/dx$) + $X$  &
	180 \mgev\ ($\stau$)\\
	&  & ll &
	all &
	$\mu$($dE/dx$+TOF) + $X$  &
	210 \mgev\ ($\stau$)\\
\hline 
$\slepton$ co-NLSP & $\ell + \gravitino$ & p &
	all & $\ell\ell\ell j \mets$ & 
	360 \mgev\ ($\schionepm$)	\\
    	& & &
	& & 160 \mgev\ ($\slepton$)	\\
    	& & d &
	all & $\ell\ell + dE/dx$ & 480 \mgev\ ($\schionepm$)	\\
    	& & &
	& & 206 \mgev\ ($\slepton$)	\\
\hline
$\stopone$ & $(c, bW) + \gravitino$ & p & 
	$\stopone\stoponeb$ & $cc\mets$ or $\ell+jets+\mets$ &
              175 \mgev\ ($\stopone$) \\
\hline 
\end{tabular}
\end{center}
\label{tab:gmsb_decay_mode}
\end{table}

The NLSP can be bino-like $\schionezero$, higgsino-like $\schionezero$,
$\stauone$, $\slepton$, or $\squark$ (likely $\stopone$) in minimal GMSB.
We here choose two scenarios, bino-like $\schionezero$ and  $\stauone$,
because the signatures define special detector
performance at CDF and D\O\ other than in Section \ref{sec:ds_sugra}.
A summary of discovery reaches for various NLSP scenarios
is given in Table~\ref{tab:gmsb_decay_mode}.

\subsubsection{Bino-like Neutralino NLSP}

For $\sqrt{F}$ greater than a few 1000 TeV, 
the $\schionezero \to \gravitino + \gamma$ decay takes place outside
a collider detector. In this case, $\schionezero$ is essentially
stable on the scale of the experiment and escapes as missing energy.
If $\sqrt{F}$ is less than that, however,
the $\schionezero$ decay takes place
within the detector.
Thus, two hard photons and large \mets\ would be observed
in all final states of pair produced SUSY particles
with cascade decays through the $\schionezero$ decay.
There is essentially no SM background.

For a representative study for $\gamma\gamma \mets + X$, 
models are chosen with
$N_{m} = 1$, $M_{m}/\Lambda = 2$, $\tanb = 2.5$, and $\mu > 0$
\cite{shw_gmsb}.
Here $\gamma\gamma$ are either prompt or displaced photons.
In the models,
the cross-section 
for $\schi\schi$ ($\schionep\schionem$ and $\schionepm\schitwozero$) production
is the largest.
Thus, the $\schionepm$ is probably the best figure of merit
for the discovery reach.

For prompt photon,
CDF and D\O\ collaborations studied the final state with different
kinematical cuts \cite{shw_gmsb_binoNLSP},
 but found similar $5\sigma$ discovery reach
(with 30 \invfb)
in $M_{\schionepm}$: 330 \mgev\ for CDF and 340 \mgev\ for D\O\
\cite{shw_gmsb}.

For a displaced photon, the D\O\  detector can reconstruct
the electromagnetic (EM) shower develpments in the EM calorimeters
to point back to the beam line to measure the distance of closest approach
($d_{ca}$) to the beam axis \cite{shw_gmsb}. 
The analysis requires at least one photon
with $d_{ca} >$ 5 cm in the final state of $\gamma_{d} jj \mets$.
Here $\gamma_{d}$ indicates a displace photon. 
For $\Lambda$ = 100 TeV 
($M_{\schionepm} \simeq 250\ \mgev$, 
$M_{\schionezero} \simeq 130\ \mgev$), 
the probability that a photon has $d_{ca} >$ 5 cm 
is about 40\% or better for $c\tau \ltsim$  100 cm.
The discovery reach in the $\schionepm$ mass ranges from 310 \mgev\
to 280 \mgev\ between $c\tau$ = 0 and 100 cm.
On the other hand, CDF will meausue the photon's arrival time
on the EM calorimeter to distinguish
from prompt photon or cosmic-ray induced photon \cite{CDF2bdet_tdr}.

\subsubsection{Stau NLSP}

The models are $N_{m} = 2$, $M_{m}/\Lambda = 3$, $\tanb = 15$, $\mu > 0$
with $\Lambda$ allowed to vary \cite{shw_gmsb}:
the $\stauone$ is lighter than $\schionezero$, but
$M_{\stauone} \sim 0.5 M_{\schionepm}$.
$M_{\sleptonR} - M_{\stauone}$ is greater than a few \mgev\ for all points.
The $\schionep\schionem$ and $\schionepm\schitwozero$
production is dominant for $M_{\schionepm}\ \ltsim\ 350\ \mgev$
and the $\stauone\stauone$ production becomes dominant for heavier
$\schionepm$.

If $\stauone$ is short-lived and decays 
in the vicinity of the production vertex,
the $\schi\schi$ production, followed by the cascade decays, will arise
in the final states of $\ell\ell\ell j\mets$ and $\ell^{\pm}\ell^{\pm}jj\mets$.
They are studied by D\O,
while CDF investigates the $\tau_h \tau_h \mets$ final state 
from $\stauone \to \tau \gravitino$. 
With 30 \invfb, both analyses have the discovery reach 
of  230 (120) \mgev\ in $M_{\schionepm\ (\stauone)}$.

For long-lived $\stauone$, we search for events containing
at least one $\mu$-like track with a large $dE/dx$ in its tracking system.
The D\O\ analysis choose the final state including two leptons 
with $M_{\ell\ell} > 50\ \mgev$ and is
sensitive for the $\schi\schi$ production.
With 30~\invfb, the discovery reaches in $M_{\schionepm}$ ($M_{\stauone}$) are
420 (210) ~\mgev.

The CDF detector includes a new time-of-flight (TOF) system.
With a timing resolution of 100 ps, we could require $4\sigma$
separation (at 400 ps), which is $\beta\gamma < 2.26$ (or $p < 235\ \mgev$).
This sould be compared to $\beta\gamma < 0.85$ using the $dE/dx$ technique
alone.
With 30~\invfb, the discovery reaches in $M_{\stauone}$ are
180~\mgev\ with $dE/dx$ and
210~\mgev\ with $dE/dx$ plus TOF.

It should be noted that the same technique ($dE/dx$, TOF) can be used \cite{shw_btmssm}
to test Anomaly-mediated SUSY Breaking (AMSB) models~\cite{amsb} 
where the $\schionepm$ is long-lived because of
a very small mass difference between $\schionepm$ and $\schionezero$.

\subsection{Summary}

The ultimate limit for the SUSY mass reach of a hadron collider, 
resulting from the distribution
function of the constituent quarks and gluons, is $\sim 25\%$ 
of its collider energy.
Thus the Tevatron approaches its limit below 450  \mgev\
(about 50\% $\sqrt{s}$)
in the discovery of the SUSY particles.
This would be similar
to the CERN Super Proton Synchrotron (S$\ppbar$S) that
approached its limit in the
discovery of the $W$ and $Z$ bosons.

\section{Detection of ${\boldmath \bsmumu}$} 

We consider now the possibility of detecting
the decay \bsmumu\ by  the CDF and D\O\ detectors at the
Tevatron in Run II \cite{Arnowitt:2002}. 
Both detectors have been upgraded with excellent tracking and
muon detector systems \cite{Anikeev:2001rk}. 
The dimuon trigger is the key to collect the 
$B_{s} \to \mu^+\mu^-$ decays.

This process is particularly  interesting
for several reasons: The SM branching ratio is quite small,
\ie,   $Br[\bsmumu]_{\rm SM} = 3.5 \times 10^{-9}$ \cite{Anikeev:2001rk}.  
The SUSY contribution
\cite{Babu:1999hn,Chankowski:uz,Bobeth:2001sq,Dedes:2001fv,Isidori:2001fv,Huang:2002dj,Mizukoshi:2002} 
has terms that grow as $\tan^{6}\beta$
and thus can become quite large  for large \tanb. 
Finally, as we shall show below, the two collider  detectors
will  be sensitive to this decay for $\tan\beta\gtsim\ 30$ in Run II.

In order to estimate the limits on $Br[B_s \to \mpmm]$ detection,  
we use the 95$\%$ C.L. limit on $Br[B_s \to \mpmm]$ published
by CDF\cite{Abe:1998ah}. 
Thus our discussion is based on the CDF detector, 
although both CDF and D\O\ detectors should have a similar
perfromance.

In the Run I analysis, CDF observed one candidate that was
consistent with $B_s \to \mpmm$ with an estimate of 
0.9 backound (BG) events
in 98 \invpb\ \cite{Abe:1998ah}. 
The primary Run-I selection variables and cut values were
$c\tau \equiv L_{xy} M_{B} / p_{T}^{\mu\mu} >100\ \mu$m,  
$I\equiv p^{\mu\mu}_T/[p^{\mu\mu}_T+\Sigma p_T]>0.75$ 
for the muon pair, and $\Delta\Phi<0.1$ rad.  
Here, $L_{xy}$  is the transverse decay length;
$p_{T}^{\mu\mu}$ 
is the transverse momentum of the dimuon system. 
$\Sigma p_T$ is the scalar sum of the transverse momenta of
all tracks, excluding the muon candidates, within a cone of 
$\Delta R\equiv\sqrt{(\Delta\eta)^2+(\Delta\phi)^2}=1$ 
around the monetum vector of the muon pair. 
The $z$ coordinate
of each track  along the beam line \cite{def} must be 
within 5 cm of the primary vertex. 
$\Delta\Phi$ is an opening azimuthal angle between 
$p_T^{\mu\mu}$ and the vector pointing from the
primary vertex to the secondary vertex 
(the reconstructed $B$-meson decay position). 
As a conservative estimate, CDF took the one event as signal
to calculate 95\% C.L. limit of signal events 
($N_1^{95\%}\equiv$ 5.06 events \cite{Abe:1998ah}) 
and had set a limit of $Br < 2.6 \times 10^{-6}$. In the analysis,  
the selection efficiency ($\eps$) for
signal events and the rejection power (${\cal R}$) for background events 
(pass a baseline selection \cite{Abe:1998ah}) are
estimated to be $\eps_{1} = 0.45$ and $\calR_{1} = 440$ by using a sample of
like-sign dimuon events ($5 < M_{\mu\mu} < 6\ \gevcc$).

In Run II, a dimuon trigger in Ref.~\cite{Anikeev:2001rk} 
will improve the acceptance for
signal events by a factor of 2.8. 
The trigger will soon be tested using the Run IIa (2 \invfb) data. 
This will allow us to modify the trigger design for
the higher luminosity expected in Run IIb (15 \invfb). 
In this paper, we assume that the
dimuon trigger can be designed by maintaing the acceptance for signal events. 
We expect to improve the acceptance for signal events by a factor of
2.8 \cite{Anikeev:2001rk}. 
If we assume the factor 2.8 to be the same for BG events, 
then we would observe 51 (386) events in  2 (15) \invfb\  with the same
cuts as in Run I. 
Therefore, CDF has to require a set of tighter cuts to obtain
the best possible upper limit.

Two types of backgrounds must be taken into account:
(i)~non-$b$ backgrounds comming from the primary vertex; 
(ii)~$b$ background events, such as the gluon-spliting
$b\bbar$ events.

One way to reduce prompt background is 
to require  a minimum decay length $L_{xy}$.
However, two tracks can appear to form a secondary vertex  
if one of two tracks originates from the primary vertex and 
the other has an impact parameter ($\delta$).
Therefore, the requirement of a minimal impact parameter 
of individual tracks can further clean up the sample. 
It has been shown for example, in the Run-I analysis for
$B^0 \to K^{0*} \mu^+ \mu^-$ events \cite{Affolder:1999eb}, 
that a tight impact parameter  cut
on significance for individual track ($\delta/\sigma_{\delta}>2$) 
significantly improve the
background rejection even with $L_{xy} > 100~\mu$m. 
One has then  $\eps \sim \eps_{1} \times 0.43$ and 
$\calR \sim \calR_{1} \times 190$.
Thus a higher track impact parameter is neccessary 
to reduce the non-$b$ backgrounds.
We would expect larger reduction with good efficiency 
even after the $L_{xy}$ cut.
The silicon vertex detector (SVX-II) will
provide us much better reduction  for the non-$b$ background than Run I. 

The most severe background in Run II will be the two muons from gluon-spliting
$b\bbar$ events. 
Since both particles are $b$ quarks, the impact parameter does not help. 
Both $b$ and $\bbar$ also  go in  the same direction,  so that cut on
$L_{xy}$ does not help either. 
However, $\Delta\Phi$ is still usefull to remove
the background events. 
Furthermore, in Run II, we can use $\Delta\Theta$ in
$r$-$z$ view since we have $z$-strips in SVX-II. 

There is some room to improve the isolation cut.  
We can form a new isolation
parameter by only using  the tracks with large impact parameter. 
This new isolation cut will work to reject the $b\bbar$ 
rather than non-$b$ background.
Furthermore, we can search for tracks with large impact parameter on the
opposite side of the dimuon candidates 
to make sure that the $b$ and $\bbar$ go
to the opposite side.

Therefore, CDF could improve the BG rejection by a factor of 200-400 with
further reduction of the signal efficiency by a factor of 2-3.  
Based on these facts, we now consider two cases 
to evaluate  Run II limits as a function of
luminosity.

In the first case (Case A), we naively assume new tighter cuts in Run II,
described above, will gain additional BG rejection power of 450 for additonal
efficiency of 0.45, or
  $ {\cal R}_2  =  450^{0.45/\eps_2}$. This gives us
\beqn
  \frac{\eps_2}{\eps_1} & = & 
	\frac{1}{1 + \log({\cal R}_2/{\cal R}_1)/\log(450)}
\label{eq:eff_vs_BGrejection}
\eeqn 
If we could optimize the BG rejection in Run IIa (2 \invfb) to be
$\calR_2 \approx 51 \calR_1$ with $\eps_2 \approx 0.61 \eps_1$ 
(from Eq.~\ref{eq:eff_vs_BGrejection}), 
then we would expect  one BG event in 2 \invfb.
Thus, with an assumption of the same size of 
the  total systematic uncertainty in
Run~II as in Run~I, we can extrapolate 
the 95\% C.L. limit to be 
$ Br < 7.7 \times 10^{-8}$ for 2 fb$^{-1}$ using $N_1^{95\%}$.

In the second case (Case B), we simply assume the Run-II background rejection
could be improved
(without loosing the signal efficiency) 
to keep the expected BG events in 2 \invfb\
at the level of Run I (\ie, 0.9 events).
If we would observe one event in 2 \invfb,
then
we could set the limits by scaling the Run-I $Br$ limit down  by the
luminosity (2000 \invpb/98 \invpb) and the acceptance by (2.8/1.0). 
Thus we obtain
$Br < 4.6 \times 10^{-8}$.
This would certainly be the optimistic scenario, but
it would be a goal of this analysis in Run IIa.
Here, the systematic uncertainty in Run II is assummed 
to be the same as in Run I. 

\begin{figure}
\begin{center}
    \epsfig{file=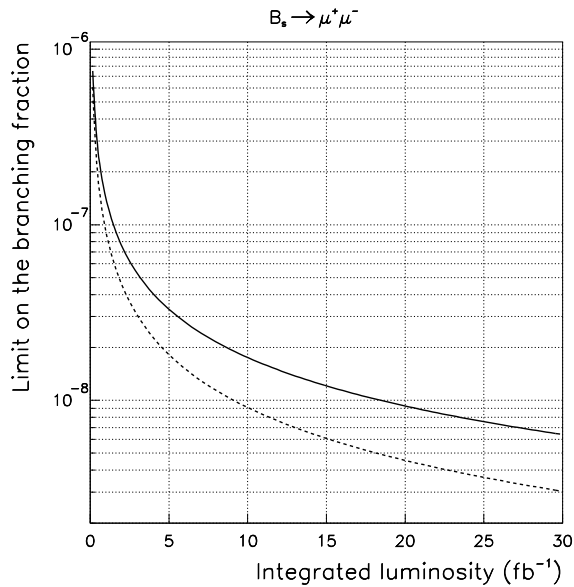,height=220pt}
\end{center}
    \caption{Illustrated 95\% C.L. limits on the branching ratio for
	$B_s \to \mpmm$ at CDF in Run II as a function of
	integrated luminosity \protect\cite{Arnowitt:2002}.
	Solid (Case A) and dashed (Case B) 
	curves are based on different assumptions
	on the signal selection efficiency and the background
	rejection power. See the text for details.}
\label{fig:bsmumu_br_limits}
\end{figure}

We repeat the same argument for different luminosities.
Figure~\ref{fig:bsmumu_br_limits} shows 
95\% C.L. limits on $Br[B_s \to \mpmm]$
at CDF in Run II as a function of integrated luminosity for Cases A and B. 
For 15 $\rm fb^{-1}$ in case A, CDF would be sensitive to
$Br>1.2\times 10^{-8}$ and the combined
CDF and D\O\ data (30 $\rm fb^{-1}$) would be sensitive
to $Br>6.5\times 10^{-9}$.

\begin{figure}
\centerline{\hbox{\epsfig{figure=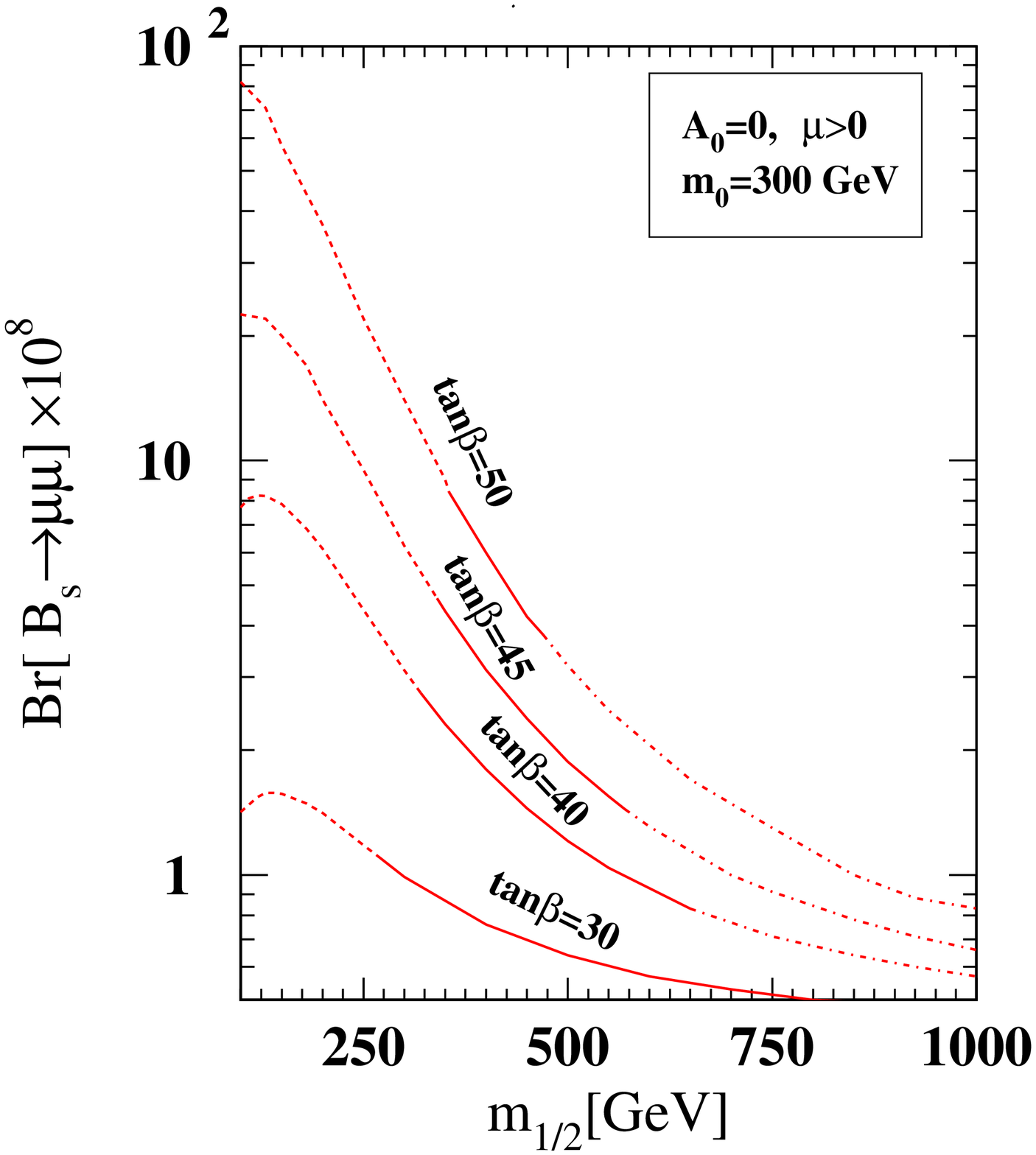,height=250pt}}
{\epsfig{figure=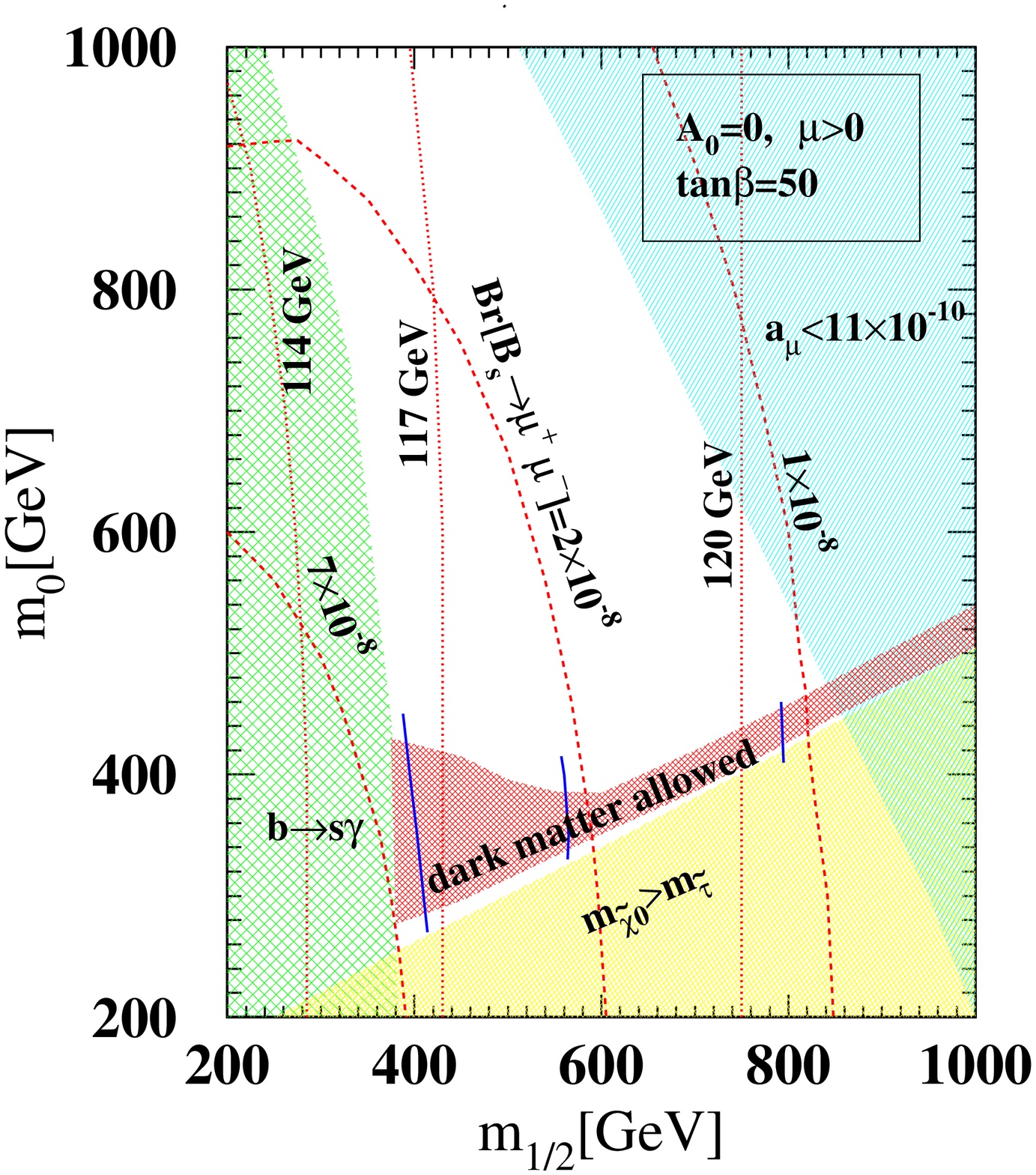,height=250pt}}}
\caption{Branching ratio for \bsmumu\
in mSUGRA models \protect\cite{Arnowitt:2002}. 
On the left plot, $Br[\bsmumu]$ is shown
as a function of $\mhalf$ for various $\tanb$ values, where
dashed and dash-dotted lines are to indicate the
models are excluded via constraints on $Br[b \to s \gamma]$ 
and $m_{\stau} > m_{\lsp}$, respectively.
On the right plot, three dashed lines from left
to right in the $\mzero$-$\mhalf$ plane indicate
$Br$ = $7\times 10^{-8}$, $2\times 10^{-8}$, $1\times 10^{-8}$,
respectively,
for $\tanb$ = 50.
The three short solid lines indicate the $\sigma_{\tilde\chi^0_1-p}$ values 
(from left:  0.05 $\times 10^{-6}$ pb,  
0.004 $\times 10^{-6}$ pb, 0.002 $\times 10^{-6}$ pb).  
The vertical dotted lines label Higgs masses.}
\label{fig:sugra1}
\end{figure}

We examine first the parameter region for the
mSUGRA model that would be  accessible to CDF or D\O\ at 
Run II with 15 \invfb\ of data. 
Figure~\ref{fig:sugra1}(left) shows the  
$Br[\bsmumu]$ as a function of \mhalf\ for \azero\ = 0, 
\mzero \,= 300 GeV. 
One sees that with a sensitivity of 
$Br[\bsmumu] > 1.2 \times 10^{-8}$
for 15 \invfb, the Tevatron Run II can 
probe the \bsmumu\  decay  for $\tanb\ > 30$.  
Further, a search for this decay would sample much higher regions of 
\mhalf\  than a direct search at Run II for SUSY particles which is restricted
to $\mhalf < 250\ \gev$  \cite{Barger:1998hp}. 
As \mzero\ increases, the branching ratio goes down.  
However, this dependence becomes less significant for large 
\mhalf, where 
\mzero\  as large as 800 GeV can be sampled for large \mhalf.

In Figure~\ref{fig:sugra1}(right)
 the contours of $Br[\bsmumu]$ are plotted in the 
\mzero-\mhalf\  plane for \tanb\ = 50, \azero\ = 0.  
Also see Figure~\ref{fig:Dedes2002_3L}.
We combine now this result
with the other  experimental constraints. 
Thus the shaded region to the left is
ruled out  by the \bsgam\     constraint, and the shaded region on the right
hand side  is disallowed if $a_\mu^{\rm SUGRA} > 11 \times 10^{-10}$.  
The narrow shaded band in the  middle is 
allowed by the dark matter constraint. 
We note that independent  of 
whether the astronomically observed dark matter is
SUSY in origin, the  dark matter allowed region for mSUGRA cannot significantly
deviate from  this shaded region, for below the narrow shaded band, the stau
would be  lighter than the neutralino (leading to charged dark matter), while
above  the band, mSUGRA would predict more neutralino dark matter than is
observed.

Using our estimate that $Br > 1.2 \times 10^{-8}$  can be observed 
with 15 \invfb,  we see that almost the entire parameter space allowed 
by the $a_\mu < 11 \times 10^{-10}$ 
constraint can be probed in Run II for \tanb\ = 50. 
Note that an observed 
$Br[\bsmumu] >  7 \times 10^{-8}$, possible with only 2 \invfb 
(see Figure~\ref{fig:bsmumu_br_limits}),  
would be sufficient to rule out the mSUGRA model
for $\tan\beta\leq 50$. 
In Figure \ref{fig:sugra1}(right)
we also show  the expected dark matter detector cross
section for Milky Way dark matter  (the short solid lines). 
For  \tanb\  = 40 and \azero\ = 0, 
we see about half the parameter space can be
scanned by the CDF detector 
(and the whole parameter space with 30 \invfb\ if a similar
analysis holds for the D\O\ detector). 
We note, further, that if \azero\ = 0,  a simultaneous measurement  of both
$Br[\bsmumu]$ and $a_\mu$ would essentially 
determine the mSUGRA  parameters, as
the \mzero\ allowed region at fixed \mhalf\
 is very narrow due to  the dark matter constraint.  
The effect of varying \azero\ is also discussed in Ref. \cite{Arnowitt:2002}.

In Figures~\ref{fig:sugra1}(right), 
we have also drawn  lines for
various light Higgs masses (vertical dotted lines). 
A measurement of $\bsmumu$,
$a_{\mu}$ and $m_h$ would then effectively determine the parameters of mSUGRA
for $\mu>0$ by requiring that they intersect with the dark matter allowed band 
at a point. 
(If no choice of parameters allowed this, mSUGRA would be ruled
out.) 
The Tevatron Run II should be able to either rule out a Higgs mass or give
evidence for its existence 
at the 3$\sigma$ level over the entire allowed mass range 
of SUSY light Higgs masses.
Alternatively, 
the LHC's determination of $m_h$ or the gluino mass (to determine
$m_{1/2}$) would fix the parameters of mSUGRA.

Thus if $a_\mu$ increases, 
the bound moves downward, encroaching further on the allowed part of the 
parameter space, and a value of 
$a_\mu \,\gtsim\ 50 \times 10^{-10}$ would eliminate 
the  mSUGRA model \cite{Arnowitt:2001be}. 
However, if $a_\mu$ decreases significantly (but is still  positive),
the mSUGRA
model would predict a heavy SUSY particle spectra  closer to the TeV region,
having significant effects on accelerator and  dark matter detection physics. 
An accurate determination of $a_\mu$  corresponds 
to a line from upper left to
lower right (or more precisely a  band when errors are included) 
running parallel to the $a_\mu < 11 \times 10^{-10}$  boundary, 
and cutting through the allowed dark
matter band which runs from  lower left to upper right. 
Thus, these two experiments are complementary 
for  determining the mSUGRA parameters.

It is also pointed out that the \bsmumu\ decay is very powerful in testing
the SUSY breaking mechanism  \cite{Baek:2002}.
If the \bsmumu\ decay is observed with $Br > 10^{-8}$,
the Tevatron could exclude 
(i)~GMSB models (with $\tanb < 40$ and lower $N_m$) and
(ii)~minimal AMSB models.

Interestingly, unlike the R parity conserving models, 
the SUSY contribution  to \bsmumu\ can occur at the tree level
in the R parity violation models.
The  $Br[\bsmumu]$ can be large for both small and large $\tan\beta$.
We, for example, consider 
lepton violating terms \cite{jang}, and so we set $\lambda^{\prime\prime}$
to zero (to prevent rapid  proton decay). 
In Ref. \cite{Arnowitt:2002}, the branching ratio
of $10^{-7}$ (feasible with 2 \invfb) could
be sensitive upto $\mhalf\ \ltsim\ 1000\ (750)\ \mgev$
for $\mzero = 300\ (500)\ \mgev$.

\section{Conclusion}

We have investigated the prospects for studying SUSY
parameter space in Run II of the Fermilab Tevatron.
Two complementary obeservables (especially in SUGRA models) 
are of great interest~\cite{Dedes:2002}:
trilepton final state as production of charginos and neutralinos
for lower \tanb\ values
and the rare  decay mode \bsmumu\ for large \tanb\ values.

Especially, the \bsmumu\ decay is an important process for SUSY searches, 
as the Standard Model prediction of the branching ratio is quite small 
($3.5 \times 10^{-9}$),  
and the SUSY contribution increases for large $\tan\beta$ as 
$\tan^{6}\beta$.  
We  find that a $Br > 1.2 \times 10^{-8}$ for 15 fb$^{-1}$ 
can be probed by each collider detector in Run II. 
(This is nearly a factor of  100 improvement over the Run I bound.) 
For the mSUGRA model, the above sensitivity implies that 
Run II  could probe a region of parameter space for $\tanb > 30$, 
a region which  could not be probed by a direct search at Run II.  
A large branching ratio, 
\ie,  $> 7 \times 10^{-8} (14 \times 10^{-8})$ would  be sufficient to 
eliminate the mSUGRA model for $\tan\beta\leq 50(55)$
if one combined the 
expectations for \bsmumu\  for mSUGRA 
with other experimental bounds on  the parameter space. 

\section*{Acknowledgments}

I wish to thank the organizers of the SUSY02 conference
for their excellent hospitality.
I should thank CDF and D\O\ collaborations to provide
valuable information and discussions.

\renewcommand{\baselinestretch}{1}


\begin{thebibliography}{99}
\bibitem{D03L:1998} 
\DzeroC, B.~Abbott \etal, 
	\PRL {\bf 80}, 1591 (1998).
	The upper limit on $\sigma \cdot Br$ is given for 
	a single trilepton mode.
	To compare with the CDF result~\protect\cite{CDF3L:1998}, 
	the D\O\ limit should be scaled up by 4.
\bibitem{def} 
We use a coordinate system where $\theta$ and $\phi$ are the polar and 
azimuthal angles, respectively, 
with respect to the proton beam direction ($z$ axis). 
The pseudorapidity ($\eta$) is defined as $-ln[\tan(\theta/2)]$. 
The transverse momentum of a particle is denoted as
$p_T=p\ \sin\theta$.
\end{thebibliography}
\end{document}